# Twist engineering of the two-dimensional magnetism in double bilayer chromium triiodide homostructures

Hongchao Xie[1,+], Xiangpeng Luo[1,+], Gaihua Ye[2,+], Zhipeng Ye[2], Haiwen Ge[3], Suk Hyun Sung[4], Emily Rennich[5], Shaohua Yan[6], Yang Fu[6], Shangjie Tian[6], Hechang Lei[6], Robert Hovden[4], Kai Sun[1], Rui He[2,*], & Liuyan Zhao[1,*]

[1] Department of Physics, University of Michigan, 450 Church Street, Ann Arbor, MI, 48109, USA
[2] Department of Electrical and Computer Engineering, Texas Tech University, 910 Boston Avenue, Lubbock, TX, 79409, USA
[3] Department of Mechanical Engineering, Texas Tech University, 2703 7th Street, Lubbock, TX 79409, USA
[4] Department of Materials Science and Engineering, University of Michigan, 2300 Hayward Street, Ann Arbor, MI, 48109, USA
[5] Department of Mechanical Engineering, University of Michigan, 2350 Hayward Street, Ann Arbor, MI, 48109, USA
[6] Department of Physics and Beijing Key Laboratory of Opto-electronic Functional Materials & Micro-Nano Devices, Renmin University of China, Beijing, 100872, China

[*] Email: rui.he@ttu.edu; lyzhao@umich.edu
[+] These authors contributed equally

**Twist engineering, or the alignment of two-dimensional (2D) crystalline layers with desired orientations, has led to tremendous success in modulating the charge degree of freedom in hetero- and homo-structures, in particular, in achieving novel correlated and topological electronic phases in moiré electronic crystals [1,2]. However, although pioneering theoretical efforts have predicted nontrivial magnetism [3,4] and magnons [5] out of twisting 2D magnets, experimental realization of twist engineering spin degree of freedom remains elusive. Here, we leverage the archetypal 2D Ising magnet chromium triiodide ($CrI_3$) to fabricate twisted double bilayer homostructures with tunable twist angles and demonstrate the successful twist engineering of 2D magnetism in them. Using linear and circular polarization-resolved Raman spectroscopy, we identify magneto-Raman signatures of a new magnetic ground state that is sharply distinct from those in natural bilayer (2L) and four-layer (4L) $CrI_3$. With careful magnetic field and twist angle dependence, we reveal that, for a very small twist angle (~ 0.5°), this emergent magnetism can be well-approximated by a weighted linear superposition of those of 2L and 4L $CI_3$ whereas, for a relatively large twist angle (~ 5°), it mostly resembles that of isolated 2L $CrI_3$. Remarkably, at an intermediate twist angle (~ 1.1°), its magnetism cannot be simply inferred from the 2L and 4L cases, because it lacks sharp spin-flip transitions that are present in 2L and 4L $CrI_3$ and features a dramatic Raman circular dichroism that is absent in natural 2L and 4L ones. Our results demonstrate the possibility of designing and controlling the spin degree of freedom in 2D magnets using twist engineering.**

Moiré superlattice forms when two vertically stacked atomic crystals are rotated with respect to each other, enabling a powerful venue to design and tailor the physical properties of 2D materials, including electronic, phononic, and magnetic ones. So far, it has achieved fruitful results in controlling the charge degree of freedom (DoF) and realizing novel quantum phenomena in both single-particle electronic states and two-particle excitonic states. Outstanding examples include the creation of flat electronic bands [1,2] that leads to various strongly correlated [6-10] and topological [11-13] phases for the former case, and the introduction of deep confinement potential that localizes excitonic states to realize moiré excitons [14,15], exciton Mott insulators [16], and quantum emitters [17] for the latter. Recently, it has also shown significant impacts on modulating the lattice DoF and induced lattice reconstructions [18,19], renormalized phonons [20], and moiré phonons [21]. In sharp contrast, the potential and power of moiré superlattices in controlling the spin DoF and engineering the magnetic properties have remained as a pristine area experimentally, despite a few pioneering theoretical predictions including noncollinear magnetism [3], topological magnetism [4], moiré magnon bands [5], and one-dimensional magnons [5].

Newly discovered layered magnets have greatly expanded the library of two-dimensional (2D) materials and provided exciting possibilities for the exploration and exploitation of the intrinsic spin DoF at the atomically thin limit [22-24]. Among the 2D magnetic atomic crystals discovered thus far, atomically thin chromium trihalides, $CrX_3$ (X = Cl, Br, and I) have attracted extensive interest as an archetype 2D magnet platform for hosting a plethora of novel 2D magnetic phenomena [25-31] and exhibiting versatile tunability with external stimuli [32-36]. In particular, it has been theoretically calculated [37] and experimentally shown [35,36] in bilayer (2L) $CrI_3$ that tuning between the monoclinic (AB') and rhombohedral (AB) stacking geometries can lead to magnetic transitions from the layered antiferromagnetic (AFM) to ferromagnetic (FM) order. Such a close relationship between structural stacking symmetries and interlayer magnetism in $CrI_3$ naturally stimulates the curiosity of investigating the magnetism in $CrI_3$ moiré magnets where the interlayer exchange coupling is periodically modulated in both sign (i.e., AFM vs FM) and strength (i.e., magnitude).

Here, we fabricate twisted double bilayer (tDB) $CrI_3$ homostructures with individual 2L $CrI_3$ to engineer the inter-2L exchange coupling and investigate the resulting moiré magnetism. Our choice of 2L $CrI_3$ as the building block is based on its significantly narrower structural and magnetic phonon linewidths than those of monolayer $CrI_3$, suggesting a much better crystalline and magnetic integrity in 2L $CrI_3$ [38,39]. Figure 1a top panel shows a false-color optical image of a typical tDB $CrI_3$ sample that was made by tearing a large piece of 2L $CrI_3$ in the bottom panel into two and then stacking them together at a controlled twist angle of $\alpha$ (see Methods). Electron diffraction (Fig. 1b) shows two sets of 1$^{st}$ and 2$^{nd}$ order Bragg peaks for the two 2L $CrI_3$ in a tDB $CrI_3$ sample with the targeted $\alpha$ of 1.0° during fabrication, from which the actual $\alpha$ was determined to be 0.9° on average with a standard deviation of 0.1° through surveying nine different locations on this sample. The match between targeted and measured values confirms our well control of the twist angle in fabrications, and the small standard deviation suggests the homogeneity of our samples. Dark field transmission electron microscopy (TEM) (Fig. 1c) displays the real-space periodic superstructures with noticeable domain formation. The bright triangles are associated with the strongly coupled regions between two 2L $CrI_3$ whereas the dark boundaries represent the decoupled boundaries. Similar reconstructing has been observed in low-twist angle graphene bilayers[19] and transition metal dichalcogenide bilayers [18].

Magneto-Raman spectroscopy can capture the interlayer magnetism in few-layer $CrI_3$, by detecting the unique static magnetism-coupled phonons that break the time-reversal symmetry and have antisymmetric Raman tensors [38-42], in addition to the conventional pure structural phonons. Figure 1d presents representative Raman spectra of tDB $CrI_3$ with the targeted $\alpha = 1.1°$ in both the crossed and parallel linear polarization channels at 10 K, featuring key Raman modes in three spectral ranges, 75 to 85 $cm^{-1}$, 95 to 120 $cm^{-1}$, and 120 to 133 $cm^{-1}$. These modes are coarsely comparable to those of few-layer $CrI_3$ among which the Raman modes in the 75 to 85 $cm^{-1}$ and 120 to 140 $cm^{-1}$ ranges particularly highlight the contribution from the magnetism-coupled-phonon scattering [38]. In this work, we focus on the 120 to 133 $cm^{-1}$ range because the Raman modes here could be related to the moiré magnetism and are of higher intensity than those in 75 to 85 $cm^{-1}$. Figure 1e zooms into the Raman spectra in the 120 to 133 $cm^{-1}$ range and includes data in both linearly crossed and parallel channels with incident polarizations at the horizontal and 45° rotated directions, i.e., $\theta_{inc} = 0°$ and 45°. Clearly, the one primary mode in the parallel channels ($U_1^t$ at 129.4 $cm^{-1}$) and the three modes in the crossed channels ($U_{2,3,4}^t$ at 129.0 $cm^{-1}$, 127.4 $cm^{-1}$, and 126.3 $cm^{-1}$) all and individually show no polarization dependence, confirming that Raman modes in the crossed (parallel) channels correspond to antisymmetric (fully symmetric) Raman tensors, thus break (preserve) the time-reversal symmetry, and correspond to the static magnetism-coupled (pure structural) phonon contribution, similar to those in natural few-layer $CrI_3$ [38]. On the other hand, the triplet lineshape of the tDB $CrI_3$ Raman spectra in the crossed channel shows a clear distinction from those of 2L and 4L $CrI_3$ with a single ($U_2^{2L}$ at 127.1 $cm^{-1}$) and two ($U_2^{4L}$ at 128.6 $cm^{-1}$ and $U_4^{4L}$ at 125.7 $cm^{-1}$) Raman modes, respectively [38], shown in Fig. 2a. Such a distinction strongly suggests the substantial difference in the magnetism between tDB $CrI_3$ and natural 2L/4L $CrI_3$.

Having established the Raman signature for the magnetism of tDB $CrI_3$ and its distinction from those of natural 2L/ 4L $CrI_3$, we proceed to examine its twist angle dependence. Figure 2a collects Raman spectra over 120 – 133 $cm^{-1}$ taken on tDB $CrI_3$ of selected $\alpha = 0.5°, 1.1°, 2.0°,$ and 5.0°, as well as 4L and 2L $CrI_3$, in both the crossed and parallel linear polarization channels at 10 K. We observe that the central mode ($U_3^t$) in the triplet in the crossed channel increases in intensity and the two modes on its side ($U_{2,4}^t$) decrease as the twist angle $\alpha$ increases, showing the trend that the magneto-Raman spectra of tDB CrI$_3$ evolves from resembling most the natural 4L CrI$_3$ at the lowest $\alpha$ to converging towards the 2L CrI$_3$ at the highest $\alpha$. This trend is expected because the inter-2L coupling strength in tDB CrI$_3$ weakens at larger $\alpha$, leading to the parallel stacked ($\alpha = 0°$) tDB CrI$_3$ relaxing to a 4L CrI$_3$ flake and the large twist angle ones are equivalent to two decoupled 2L CrI$_3$ films. This intensity evolution is further quantitatively summarized in Fig. 2c where the relative intensity ratio $I_{U_3^t}/(I_{U_2^t} + I_{U_4^t})$ is plotted against $\alpha$ for tDB CrI$_3$ and compared with those of 4L and 2L CrI$_3$, showing a monotonous enhancement with increasing $\alpha$ and further confirming the corresponding reduction of inter-2L coupling at larger $\alpha$. Equally informative is the frequency shift of the Raman modes which is shown in Fig. 2b. At low twist angles (*e.g.*, $\alpha = 0.5°$, $1.1°$), the frequencies of $U_2^t$ and $U_4^t$ in tDB CrI$_3$ match with those of $U_2^{4L}$ and $U_4^{4L}$ in 4L CrI$_3$, respectively, whereas the frequency of $U_3^t$ is close to that of $U_2^{2L}$ in 2L CrI$_3$ which appears in the crossed channel (the same as $U_3^t$) and that of $U_3^{4L}$ in 4L CrI$_3$ which is however absent in the crossed channel (in contrast to $U_3^t$). As the twist angle increases, the frequencies of $U_2^t$ and $U_4^t$ blueshift towards their high-frequency neighbors $U_1^t$ and $U_3^t$. Eventually, at large twist angles (*e.g.*, $\alpha = 5°$), the frequencies of $U_{2,4}^t$ become nearly indistinguishable from $U_{1,3}^t$ and approach those of $U_{1,2}^{2L}$ in 2L CrI$_3$. The correspondence of mode frequencies between the fabricated tDB CrI$_3$ and the natural 4L/2L CrI$_3$ reveals that magnetism-coupled-phonon contributions for $U_2^t$ and $U_4^t$ arise from regions with strong inter-2L coupling, resembling the 4L-like case, whereas that for $U_3^t$ is dominated by the decoupled regions, mimicking the 2L-like case. The twist angle dependencies of mode frequencies and the relative intensity ratio both confirm the reduction of inter-2L coupling and the suppression of strongly coupled 2L-2L regions in tDB CrI$_3$ at larger $\alpha$.

Out-of-plane magnetic field ($B_\perp$) is known to introduce sharp spin-flip transitions in few-layer CrI$_3$, which is nicely captured in the $B_\perp$ dependence of Raman spectra of the magnetism-coupled phonon scattering [38]. We now proceed to more in-depth investigations of the engineered magnetism in tDB CrI$_3$ by performing its magnetic field dependencies for selected twist angles and comparing them across one another and with natural 4L/2L CrI$_3$. Figures 3a-e present the false color maps of the $B_\perp$ dependent magneto-Raman spectra for 4L CrI$_3$, tDB CrI$_3$ at selected twist angles of $\alpha = 0.5°$, $1.1°$, and $5.0°$, and 2L CrI$_3$, respectively, in the linearly crossed channel at 10 K, and Figures 3f-j summarize their corresponding $B_\perp$ dependencies of the fitted mode intensities of the magnetism-coupled phonons. We first briefly summarize the evolutions of $U_i^{4L/2L}$, with $i = 1 - 4$ for 4L and $i = 1 - 2$ for 2L CrI$_3$, upon increasing $B_\perp$ as a reference for understanding those of $U_{1-4}^t$ for tDB CrI$_3$. For 4L CrI$_3$, $U_1^{4L}$ and $U_3^{4L}$ emerge abruptly at $B_{c1} = 0.7$ T and then sharply jump upwards and downwards, respectively, at $B_{c2} = 1.6$ T, whereas $U_2^{4L}$ and $U_4^{4L}$ both experience two consecutive steep drops at $B_{c1}$ and $B_{c2}$, with plateaus of constant intensities outside of $B_{c1}$ and $B_{c2}$, for which $B_{c1}$ and $B_{c2}$ correspond to two first-order spin flip transitions for the layered magnetism transiting from ↑↓↑↓ (layered AFM) first to ↑↓↑↑ and then to ↑↑↑↑ (fully spin polarized FM), respectively [25], with ↑ (↓) for out-of-plane magnetic moment aligning up (down). For 2L CrI$_3$, $U_1^{2L}$ appears and $U_2^{2L}$ disappears concurrently at $B_c$ = 0.7 T with $B_c$ for the spin flip transition from ↑↓ to ↑↑ [25].

We now turn to the magnetic field dependencies of $U_{1-4}^t$ in tDB CrI$_3$ at very small ($\alpha$ = 0.5°) and relatively large ($\alpha$ = 5°) twist angles, both of which can be well described by a simple weighted linear superposition of those of 4L and 2L CrI$_3$. Similar to the 4L case (Figs. 3a and 3f), the 0.5° tDB CrI$_3$ features two transitions at $B_{c1}$ = 0.7 T and $B_{c2}$ = 1.6 T with jumps/drops in $U_{1-4}^t$ mode intensities near $B_{c1}$ and $B_{c2}$ of the same trends as those in 4L CrI$_3$ and plateaus of constant intensities otherwise (Figs. 3b and 3g). Different from the 4L case, $U_3^t$ of the 0.5° tDB CrI$_3$ is present even below $B_{c1}$, and the transitions are slightly broadened. As discussed in Fig. 2, $U_2^t$ and $U_4^t$ only originate from regions with strong coupling between the two bilayers in tDB CrI$_3$, whereas $U_1^t$ and $U_3^t$ match in frequency with modes in both coupled and decoupled regions. Thus, the behaviors of $U_{1-4}^t$ in the 0.5° tDB CrI$_3$ can be well accounted for by a weighted add-up of the $B_\perp$ dependencies of $U_{1-4}^{4L}$ and $U_{1-2}^{2L}$ that correspond to the contributions from the strongly coupled and the decoupled 2L-2L regions, respectively. On the other hand, the 5.0° tDB CrI$_3$ in Figs. 3d and 3i nearly replicates the 2L CrI$_3$ results in Figs. 3e and 3j, except a minor remnant in $U_3^t$ above $B_c$ = 0.7 T. This observation confirms that the 5.0° tDB CrI$_3$ is primarily regarded as two decoupled 2L CrI$_3$ with only a tiny fraction of coupled regions. Therefore, the magnetism for both the 0.5° and 5.0° tDB CrI$_3$ can be well described by the combinations of those of 4L and 2L CrI$_3$, which makes sense because at very low twist angles (e.g., 0.5°), the structure of tDB CrI$_3$ relaxes to maximize the natural 4L-like regions whereas at large twist angles (e.g., 5.0°), the inter-2L coupling in tDB CrI$_3$ is so significantly reduced to approach two decoupled 2L CrI$_3$, similar to many other van der Waals homo- and hetero-structures.

Interestingly, in stark contrast to the 0.5° and 5.0° tDB CrI$_3$, the magnetism in the 1.1° tDB CrI$_3$ can no longer be understood simply from the combination of 4L and 2L CrI$_3$ because its $B_\perp$ dependence features two key behaviors that are not at all present in those of 4L and 2L CrI$_3$ (Figs. 3c and 3h). First, around $B_{c1}$ ≈ 0.6 T, the intensities of $U_2^t$ and $U_4^t$ show a dramatic spike feature whereas $U_3^t$ exhibits a sharp dip, instead of the step-like jumps in 4L/2L CrI$_3$. Second, between $B_{c1}$ and $B_{c2}$ ≈ 1.5 T, $U_1^t$ increases gradually until the saturation at $B_{c2}$ while $U_3^t$ and $U_4^t$ decrease slowly towards a finite intensity and zero, respectively, rather than the plateaus of constant intensities in 4L/2L CrI$_3$. These two clear discrepancies between the magnetic field dependencies of the 1.1° tDB and 4L/2L CrI$_3$ strongly suggest the emergence of a fundamentally new magnetic phase in the tDB CrI$_3$ of intermediate twist angles, and therefore the distinction from magnetism in both small and large twist angle tDB CrI$_3$.

The anomalous behaviors of the 1.1° tDB CrI$_3$ above are further supported by the observation of large Raman circular dichroism at 0 T which is absolutely zero in 4L and 2L CrI$_3$ [38]. Figures 4a and 4b show the raw spectra of the 1.1° tDB CrI$_3$ taken at 0 T in both the LL and RR channels and their corresponding Lorentzian fitting profiles for individual modes, where LL (RR) refers to the incident and scattered light with left- (right-) handed circular polarizations. Here, we note that we could only resolve three ($U^t_{1,3,4}$) out of the four Raman modes because $U^t_2$ is overwhelmed by the spectrally closest and much stronger $U^t_1$ in the same co-circularly polarized channels. Clearly, $U^t_1$ and $U^t_3$ show substantial differences between the LL and RR channels with opposite relative intensities, whereas $U^t_4$ is almost helicity independent. Figure 4c presents the $B_\perp$ dependencies of the fitted intensities of $U^t_{1,3,4}$ in both co-circularly polarized channels, where the field is swept from + 2 T to – 2 T and then back to + 2 T. The two key features highlighted in Fig. 3h, the sharp dip/peak around $B_{c1}$ for $U^t_3/U^t_4$ and the gradual evolution between $B_{c1}$ and $B_{c2}$ for all modes, are nicely reproduced in Fig. 4c. In addition, a third feature noticed is that $U^t_1$ and $U^t_3$ exhibit notable hysteresis loops below $B_{c1}$, consistent with their substantial Raman circular dichroism even at 0 T.

Let us point out two important facts before establishing the understanding on these three key features in the 1.1° tDB CrI$_3$. First, recent DFT calculations show that both monoclinic and rhombohedral stackings between layers are structurally favorable with nearly identical elastic energies, yet they correspond to distinct interlayer AFM and FM exchange coupling, respectively [37]. In a moiré supercell of the 1.1° tDB CrI$_3$, both monoclinic and rhombohedral stacking regions take up appreciable amount of areas and feature the strong but opposite magnetic coupling across the 2L-2L interface, leading to the competition and frustration of spins at the boundaries between the monoclinic and rhombohedral stacking regions. Second, our polar magneto-Raman geometry is only sensitive to the out-of-plane components of spins. Although our probe and analysis do not directly account for the in-plane components of spins which are likely to develop from the aforementioned frustrations [3,4], our results on the out-of-plane magnetic orders in tDB CrI$_3$, at the same time, are compatible with the presence of in-plane spin components.

Therefore, we propose the following model for the magnetism in the 1.1° tDB CrI$_3$, where the strongly coupled 2L-2L regions have two types, the monoclinic AFM and the rhombohedral FM inter-2L coupling, and the optimized out-of-plane magnetic order features ↑↓↑↓ for the monoclinic regions and ↑↑↑↓ for the rhombohedral regions. As shown in Supplemental Information S1, based on parameters obtained from DFT calculations, this spin configuration has the lowest magnetic energy for intermediate twist angles. Because the ↑↑↑↓ state generates a net magnetization and contributes an effective out-of-plane magnetic field, which causes the re-orientation of the spins in the outermost layer to require a lower external $B_\perp$ in the rhombohedral regions than that in the monoclinic regions, i.e., $B^R_{c1} < B^M_{c1}$ with R(M) denoting the rhombohedral(monoclinic) stacking. Therefore, as a function of $B_\perp$, the magnetic state of the 1.1° tDB CrI$_3$ undergoes three critical transitions at $B_{c1}^R$ = 0.5 T, $B_{c1}^M$ = 0.7 T, and $B_{c2}$ = 1.5 T, corresponding to the spin reorientation in the outermost layer in the rhombohedral and monoclinic regions, and the interior layer in the monoclinic regions, as depicted in Fig. 4d. We now can understand the three key features in the 1.1° tDB CrI$_3$. First, the dip/peak around 0.6 T for $U_3^t/U_4^t$ (Figure 3h) in fact corresponds to the narrow field range between $B_{c1}^R$ and $B_{c1}^M$, where the spin state of ↑↑↑↑ in the rhombohedral regions provides the least magnetism-coupled phonon contribution for $U_{3,4}^t$ modes (see Supplemental Information S2). Second, the gradual $B_\perp$ dependencies of all modes between $B_{c1}^M$ and $B_{c2}$ are then due to spin flop towards the $B_\perp$ direction, either by the expansion of the polarized area or by the increase in the out-of-plane components. Third, the Raman circular dichroism of $U_{1,3}^t$ at 0 T is from the interference between the individual pure structural and magnetism-coupled phonon contributions, both of which are present for the ↑↑↑↓ state in the rhombohedral regions at 0 T. Thus, the hysteresis of $U_{1,3}^t$ below $B_{c1}^R$ is naturally explained.

In summary, we have successfully engineered the 2D magnetism in tDB CrI$_3$ that exhibits a clear twist angle dependence. Although different from that of 4L/2L CrI$_3$, the magnetism in the very small and large twist angle tDB CrI$_3$ are dominated by the superpositions between 4L- and 2L-like regions, whereas that in the intermediate twist angle features fundamentally new magnetic orders that are absent in 4L/2L CrI$_3$. Our results open up experimental pathways for engineering and understanding 2D moiré magnetism, among which we highlight two immediate opportunities. One is the exploitation of in-plane spin component sensitive experimental techniques, for example, nonlinear optical spectroscopy [43-45], to develop a comprehensive understanding of moiré magnets where noncollinear spin textures with in-plane components often happen. The other is the exploration of moiré magnets for which the interlayer exchange coupling dominates over the intralayer exchange isotropy in the composing 2D magnets, such as 2D XY-type magnets [46,47], so that the periodic moiré interlayer exchange interaction is the leading order magnetic energy scale.

**Methods**

**Growth of CrI$_3$ single crystals**    Single crystals of CrI$_3$ were grown by the chemical vapor transport method. Chromium power (99.99% purity) and iodine flakes (99.999% purity) in a 1:3 molar ratio were put into a silicon tube with a length of 200 mm and an inner diameter of 14 mm. The tube was pumped down to 0.01 Pa and sealed under vacuum, and then placed in a two-zone horizontal tube furnace whose two zones were raised up slowly to 903 K and 823 K for 2 days and then held for another 7 days. Shiny, black, plate-like crystals with lateral dimensions of up to several millimeters were obtained from this growth procedure.

**Fabrications of 2D CrI$_3$ and tDB CrI$_3$**    Atomically thin 2D CrI$_3$ flakes were exfoliated in a nitrogen-filled glovebox, and their thickness was first determined by the optical color contrast to select natural 4L and 2L CrI$_3$ and further confirmed using Raman spectroscopy at 10 K. Using a polymer-stamping technique inside the glovebox, large-size (lateral dimensions greater than 10 μm) 2L CrI$_3$ flakes were torn into two parts with similar sizes, one of which was rotated by a well-controlled rotation micrometer for targeted twist angles and then brought down to stack with the remaining half. Both 4L/2L and tDB CrI$_3$ samples were sandwiched between two few-layer hBN flakes to avoid surface reactions with oxygen and moisture in the ambient environment after taking out from the glovebox. The samples for magneto-Raman spectroscopy measurements were placed onto the SiO$_2$/Si substrate, and those for TEM measurements were transferred onto to TEM grids.

**Micro-Raman spectroscopy**    Micro-Raman spectroscopy measurements were carried out using a 632.81 nm excitation laser with a full width half maximum (FWHM) of 0.85 cm$^{-1}$, on the resonance with the charge transfer and Cr$^{3+}$ + $^4A_2$ to $^4A_1$ transitions of CrI$_3$ in order to increase the Raman sensitivity. The laser beam on the sample site was focused down to ~ 3 μm FWHM in diameter using a 40× transmissive objective, and the laser power was kept at about 80 μW, in order to minimize the local heating effect during measurements. Backscattering geometry was used, where the scattered light was dispersed by a Horiba LabRAM HR Evolution Raman microscope (1800 grooves/mm grating) from Horiba Scientific and detected by a thermoelectric cooled CCD camera from Horiba Scientific. A commercial variable temperature (< 10 K – 325 K), closed cycle, microscopy cryostat from Cryo Industries of America, Inc was interfaced with the Raman microscope. A commercial cryogen free room-temperature-bore (2'' in diameter and 6.88'' long) superconducting magnet from Cryo Industries of America, Inc was used to achieve the variable out-of-plane magnetic field from 0 T to 2 T. The cryostat cold finger, on which the samples were mounted, was inserted into the center of the room-temperature-bore of the magnet. The linearly polarized magnto-Raman measurements were calibrated by suppressing the Rayleigh line in the linearly crossed channel, so as to overcome the Faraday artifact that is caused by the stray magnetic fields passing through the objective. The circularly polarized magneto-Raman measurements were not affected by this Faraday effect, and therefore no corrections were applied.

**Transmission Electron Microscopy** Crystallographic orientations of the two composing 2L $CrI_3$ flakes in a tDB $CrI_3$ homostructure were identified by selected area electron diffraction measurements on Thermo Fisher Talos operated at 200 keV, equipped with Gatan OneView camera. Each Bragg peak was fitted with a 2D Gaussian to quantify the twist angle in reciprocal space. A total of nine locations were surveyed for the homostructure to develop a statistical confidence in assigning the twist angle and its standard deviation. Figure 1c was generated by averaging DF-TEM images from three $5^{th}$ order Bragg peaks, spaced 120° apart to remove anisotropy.

**Acknowledgement**

L. Zhao acknowledges the support by NSF CAREER Grant No. DMR-174774 and AFOSR YIP Grant No. FA9550-21-1-0065. R. He acknowledges the support by NSF CAREER Grant No. DMR-1760668. K. Sun acknowledges the support by NSF Grant No. NSF-EFMA-1741618. R. Hovden acknowledges the support from W. M. Keck Foundation. This work made use of facilities at Michigan Center for Materials Characterization. H. Lei acknowledges support by the National Key R&D Program of China (Grant No. 2018YFE0202600, 2016YFA0300504), the National Natural Science Foundation of China (No. 11774423, and 11822412), the Beijing Natural Science Foundation (Grant No. Z200005), and the Fundamental Research Funds for the Central Universities and Research Funds of Renmin University of China (RUC) (Grant No. 18XNLG14 and 19XNLG17, 20XNH062).

**Author contributions**

L. Zhao, H. Xie, and X. Luo conceived the idea and initiated this project; H. Xie fabricated the 4L and 2L $CrI_3$ and tDB $CrI_3$ homostructures; H. Xie, X. Luo, G. Ye, Z. Ye, and H. Ge carried out the Raman experiments under the supervision of L. Zhao and R. He; S. H. Sung, E. Rennich, and R. Hovden performed the TEM characterizations; S. Yan, Y. Fu, S. Tian, and H. Lei grew the vdW $CrI_3$ bulk single crystals; K. Sun performed the theoretical calculations; H. Xie, X. Luo, R. He, and L. Zhao analyzed the data and wrote the manuscript; all authors participated in the discussion of the results.

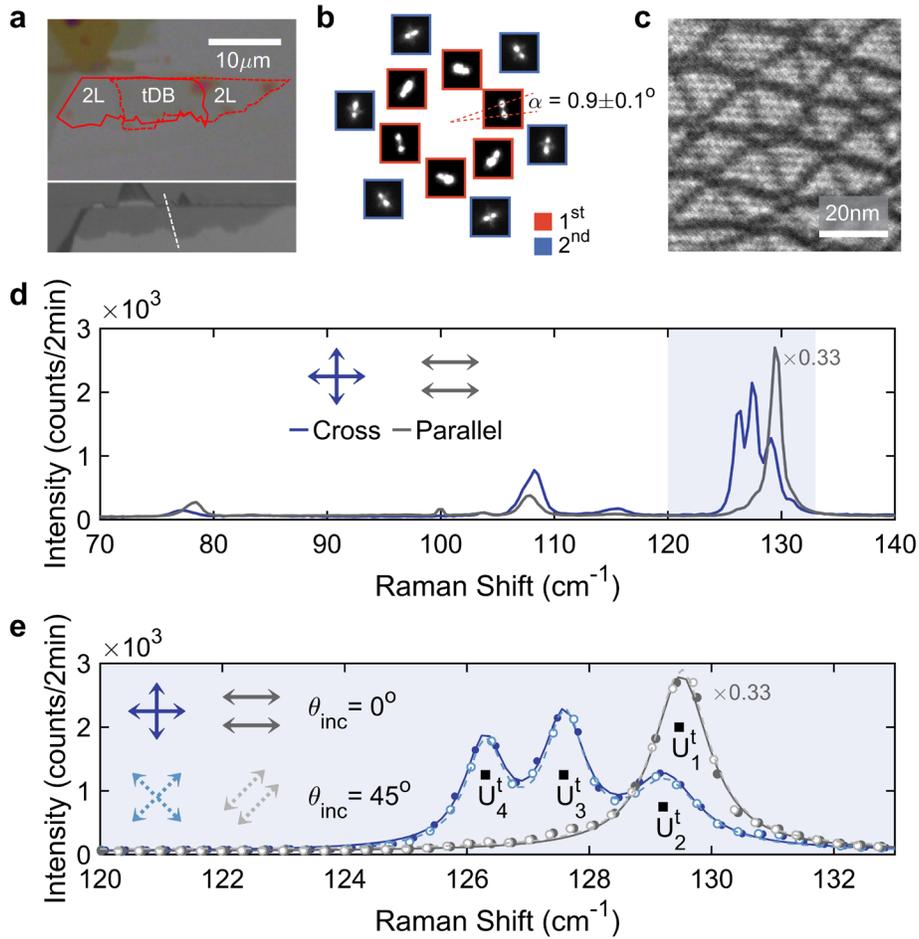

**Figure 1. Sample fabrication, TEM and magneto-Raman characterizations of tDB CrI$_3$. a.** False-colored optical images for a tDB CrI$_3$ homostructure (top panel) with two composing 2L CrI$_3$ outlined with red lines and its original large-size 2L CrI$_3$ (bottom panel) with the tearing boundary marked with a white dashed line. **b.** Electron diffraction patterns of 1$^{st}$ and 2$^{nd}$ order Bragg peaks for a tDB CrI$_3$ with a targeted twist angle α of 1.0°, showing the measured value of α = 0.9° ± 0.1° based on the survey over nine areas in this tDB CrI$_3$. **c.** Averaged DF-TEM real space image from three 5$^{th}$ order Bragg peaks showing the hexagonal superlattice network in the 0.9° tDB CrI$_3$ in **b**. **d.** Full-range Raman spectra taken on a 1.1° tDB CrI$_3$ sample in both the crossed (blue) and parallel (gray) linear polarization channels at 10 K. The parallel channel spectrum is scaled by a factor of 0.33 for a better visual comparison with the crossed channel one. **e.** Zoom-in of Raman spectra over the 120 – 133 cm$^{-1}$ frequency range taken on the same 1.1° tDB CrI$_3$ sample as in **d** in both the parallel (blue) and crossed (gray) channels at two incident polarizations of $\theta_{inc}$ = 0° (filled circles and solid lines) and 45° (open circles and dashed lines) at 10 K. The parallel channel spectra are scaled by a factor of 0.33.

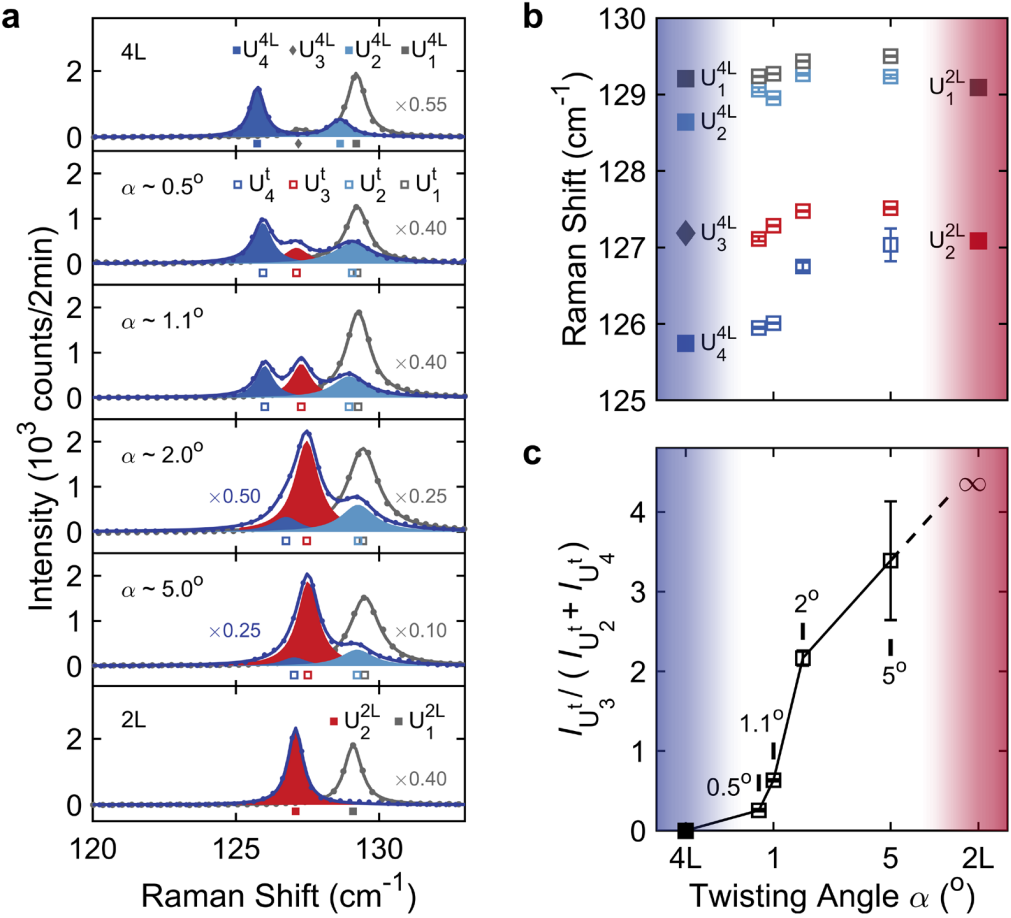

**Figure 2. Twist angle dependence of the magneto-Raman spectra of tDB CrI$_3$. a.** Raman spectra taken on 4L CrI$_3$, tDB CrI$_3$ homostructures with targeted twist angles of α = 0.5º, 1.1º, 2.0º, and 5.0º, and 2L CrI$_3$ in both the crossed (blue dots and lines) and parallel (gray) channels at 10 K. The Raman spectra are scaled by factors labeled in individual panels. The Raman modes are also marked at their corresponding frequencies, with $U^{4L}_{1,2,3,4}$, $U^{t}_{1,2,3,4}$, and $U^{2L}_{1,2}$ for 4L, tDB, and 2L CrI$_3$. The colored Lorentzian profiles highlight individual modes appearing in the crossed channel, with red (blue) for the central (two side) mode(s) in the triplet spectra of tDB CrI$_3$ and the mode(s) of 2L (4L) CrI$_3$. **b.** Plot of the fitted Raman frequencies of individual modes in every spectrum shown in **a** as a function of the twist angle. **c.** Plot of the intensity ratio $I_{U^{t}_3}/(I_{U^{t}_2}+I_{U^{t}_4})$ as a function of the twist angle. Error bars correspond to one standard error in fitting the Raman spectra.

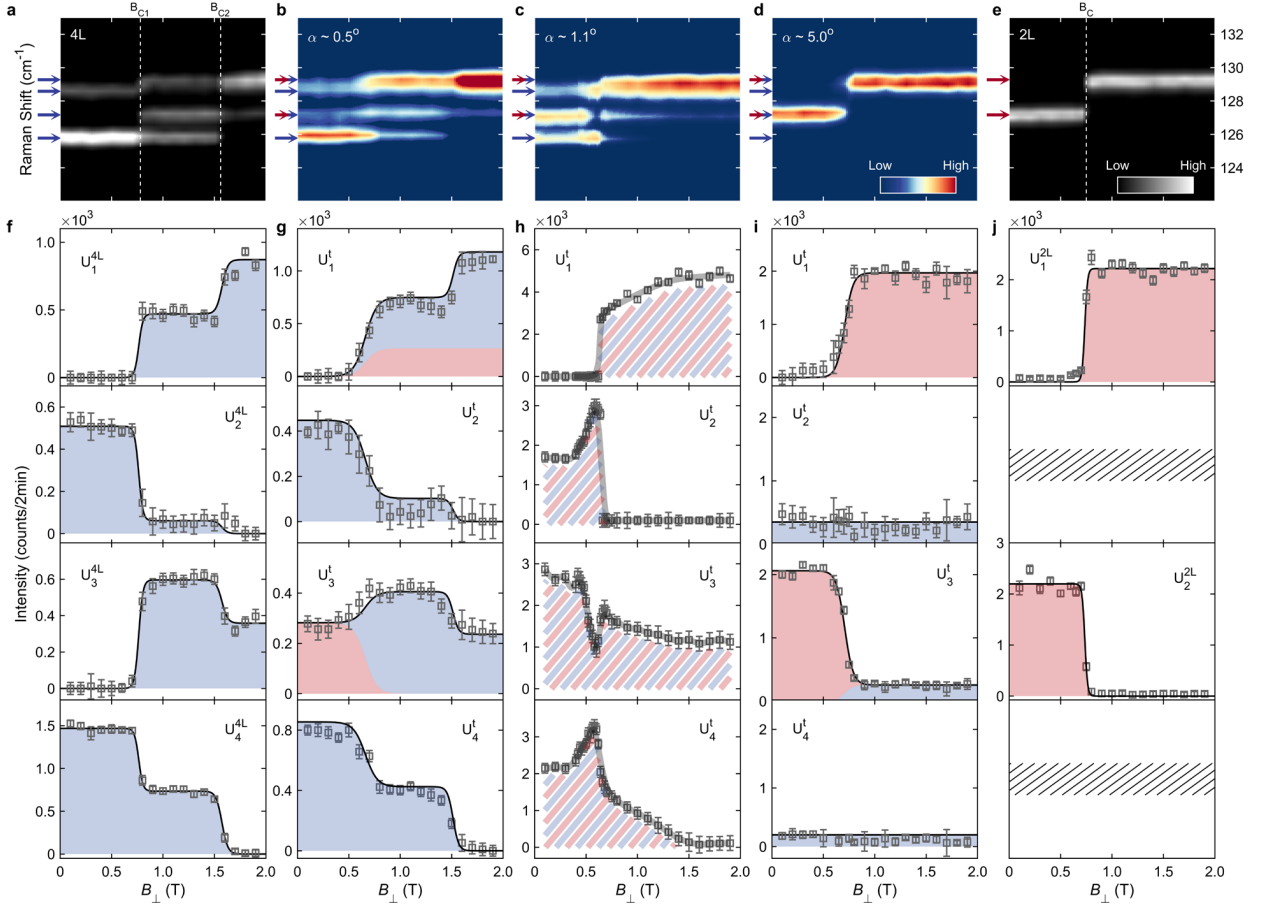

**Figure 3. Magnetic field dependence of the magneto-Raman spectra of tDB CrI$_3$ at selected twist angles. a-e.** False-colored maps of the $B_\perp$ dependent Raman spectra taken on 4L CrI$_3$, tDB CrI$_3$ homostructures with targeted twist angles α = 0.5°, 1.1°, and 5.0°, and 2L CrI$_3$, respectively, in the crossed linear polarization channel at 10 K. The blue arrows mark the frequencies of $U^{4L}_{1,2,3,4}$ for 4L CrI$_3$, and the red ones are for those of $U^{2L}_{1,2}$ for 2L CrI$_3$. **f-j.** Plots of the fitted mode intensities as a function of $B_\perp$ for the samples in **a-e**. Open squares are for the fitted values. Thin black lines and solid shaded areas are for the fits to the established model of the magnetism-coupled phonon scattering for the 4L (**f**) and 2L (**j**) CrI$_3$, and to the proposed weighted linear superpositions of 4L and 2L contributions for the very small (0.5°, **g**) and large (5.0°, **j**) twist angle tDB CrI$_3$. Thick gray lines and the striped areas in **h** are guide to the eye. Error bars correspond to one standard error in fitting the Raman spectra.

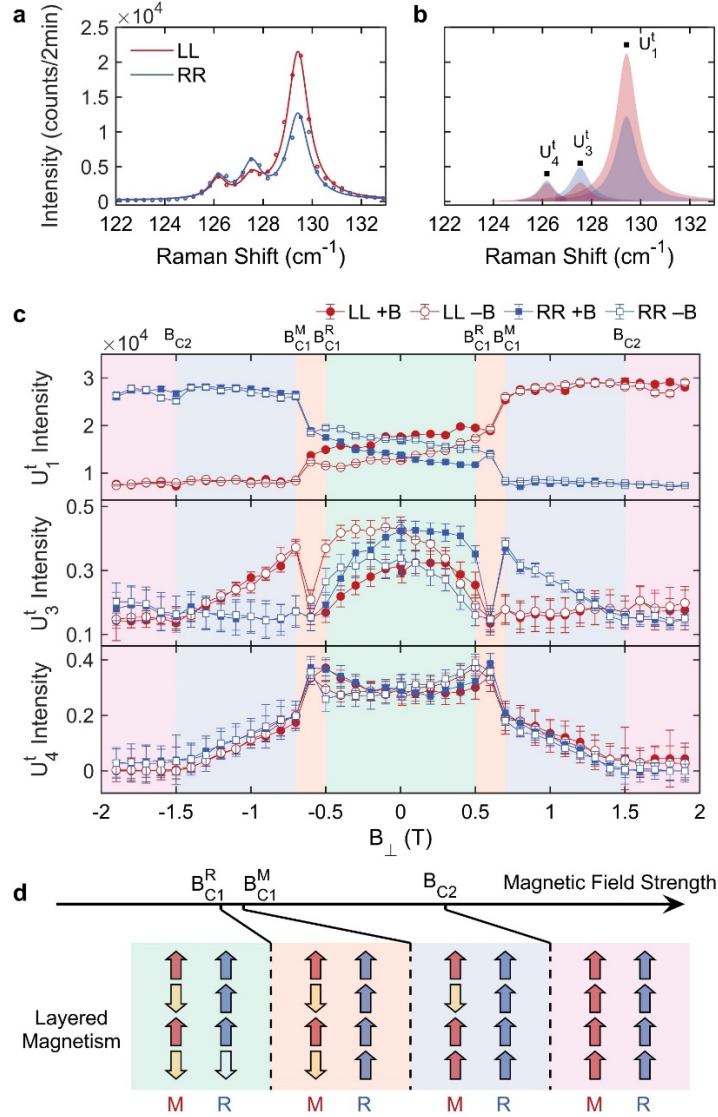

**Figure 4. Magnetic circular dichroism for the 1.1° twist angle tDB CrI$_3$. a.** Raman spectra taken in the LL (red) and RR (blue) co-circularly polarized channels at 10 K on the 1.1° tDB CrI$_3$. **b.** Fitted Lorentzian profiles for individual modes ($U^t_{1,3,4}$) in the Raman spectra in **a**. **c.** Plots of fitted mode intensities for $U^t_{1,3,4}$ as a function of $B_\perp$ that is swept from + 2 T to – 2 T (decreasing B, open circle/square for the LL/RR channel) and then to + 2 T (increasing B, solid circle/square for the LL/RR channel). Three critical magnetic fields, $B^R_{c1}$, $B^M_{c1}$, and $B_{c2}$, are marked on both the upwards and downwards magnetic fields, and four magnetic field ranges, below $B^R_{c1}$, $B^R_{c1}$ to $B^M_{c1}$, $B^M_{c1}$ to $B_{c2}$, and above $B_{c2}$ are shaded, respectively, with light green, orange, blue, and red. **d.** Sketches of the out-of-plane layered magnetism in the two types of strongly coupled 2L-2L regions of a moiré supercell, monoclinic (M) and rhombohedral (R) ones, evolving as the $B_\perp$ increases across $B^R_{c1}$, $B^M_{c1}$, and $B_{c2}$.

Supplemental Information for

# Twist engineering of the two-dimensional magnetism in double bilayer chromium triiodide homostructures

Hongchao Xie[1,+], Xiangpeng Luo[1,+], Gaihua Ye[2,+], Zhipeng Ye[2], Haiwen Ge[3], Suk Hyun Sung[4], Emily Rennich[5], Shaohua Yan[6], Yang Fu[6], Shangjie Tian[6], Hechang Lei[6], Robert Hovden[4], Kai Sun[1], Rui He[2,*], & Liuyan Zhao[1,*]

[1] Department of Physics, University of Michigan, 450 Church Street, Ann Arbor, MI, 48109, USA
[2] Department of Electrical and Computer Engineering, Texas Tech University, 910 Boston Avenue, Lubbock, TX, 79409, USA
[3] Department of Mechanical Engineering, Texas Tech University, 2703 7th Street, Lubbock, TX 79409, USA
[4] Department of Materials Science and Engineering, University of Michigan, 2300 Hayward Street, Ann Arbor, MI, 48109, USA
[5] Department of Mechanical Engineering, University of Michigan, 2350 Hayward Street, Ann Arbor, MI, 48109, USA
[6] Department of Physics and Beijing Key Laboratory of Opto-electronic Functional Materials & Micro-Nano Devices, Renmin University of China, Beijing, 100872, China

[*] Email: rui.he@ttu.edu; lyzhao@umich.edu
[+] These authors contributed equally

**Table of Contents**

S1.   Theoretical calculations for determining the out-of-plane magnetic order in tDB $CrI_3$

S2.   Calculations of the magnetism-coupled phonon contributions in tDB $CrI_3$

## S1. Theoretical calculations for determining the out-of-plane magnetic order in tDB CrI$_3$

**Background preparations**

Because the polar magneto-Raman geometry (i.e., normal incidence and backscattering geometry with the out-of-plane magnetic field) used in the current experiment, one could only access the out-of-plane components of spins due to the optical selection rule. Therefore, in our model analysis below, we will only focus on the out-of-plane components of spins. In principle, the frustration among spins in moiré magnets could result in spin canting off the out-of-plane direction even in Ising-type magnets [1]. While we do not explicitly include possible in-plane components in our model, it is worthwhile pointing out that the main conclusions from the model calculations below are not sensitive to the presence or absence of spin canting off the out-of-plane direction and are compatible with the noncollinear magnetism.

In a moiré superlattice, the periodic modulations of the interlayer stacking geometry lead to the magnetic domain structures in twisted CrI$_3$ [2]. While the structural stacking pattern and the resulting magnetic domain structure are sophisticated, it has been shown theoretically that two types of stacking geometries, AB (i.e., rhombohedral) and AB' (monoclinic) stacking, have the lowest elastic energy [3], and therefore, shall be the dominant ones. Instead of trying to capture all microscopic details about the stacking pattern and the lattice reconstruction, in our theory model, we will take a simplified setup and only focus on the most relevant and dominant parts: the monoclinic AB' and the rhombohedral AB stacking, while omitting the other high-energy stacking configurations. In reality, these high-energy stacking area may arise and fill up the regions other than the monoclinic and rhombohedral areas, especially at twist angles greater than ~ 1.0° [2]. But, the omission of them is sufficient to the leading order approximation.

In the twisted double bilayer (tDB) CrI$_3$ samples studied in our current work, the interlayer exchange coupling within individual bilayer (2L) CrI$_3$ remains antiferromagnetic (AFM), just the same as that in regular isolated 2L CrI$_3$, for which we take the value from the literature, $J_1 = 0.04$ meV/$\mu_B^2$ [2,3]. The coupling between the two 2L CrI$_3$ (i.e., at the interface between the two 2L CrI$_3$) requires the consideration of two types of structural stacking: monoclinic AB' stacking that corresponds to interlayer AFM exchange coupling and rhombohedral AB stacking that has interlayer ferromagnetic (FM) exchange coupling. We again adopt the values from literature [2,3], $J_1 = 0.04$ meV/$\mu_B^2$ for AFM and $J_2 = -0.6$ meV/$\mu_B^2$ for FM. Please note that, according to the first-principles calculations [2,3], $|J_2|$ is significantly greater than $|J_1|$. As it will be shown later, this fact is crucial for the development of net magnetization in the tDB CrI$_3$ with intermediate twist angles. Finally, the intralayer exchange coupling is known to be FM ($J < 0$) and the first-principles calculations give $J = -2.2$ meV/$\mu_B^2$.

**Theoretical modeling in tDB CrI$_3$**

For the monoclinic inter-bilayer stacked regions, the interlayer coupling between any two neighboring layers are all AFM with $J_1 = 0.04$ meV/$\mu_B^2$, and therefore its magnetism favors the interlayer AFM order: ↑↓↑↓ or ↓↑↓↑ from the top to the bottom layer. In contrast, for the rhombohedral inter-bilayer stacked regions, the preferred spin alignment is ↑↓↓↑ or ↓↑↑↓ because of the FM exchange coupling at the bilayer-bilayer interface. *It is important to note that for both stacking geometries, all these energetically favored spin configurations have zero net magnetization.* In the other word, neither of the stacking favors the development of a net magnetization that is observed in our experiment: *the large Raman circular dichroism for the 1.1° twisted tDB CrI$_3$ at 0 T.* As we shall see below, to achieve the experimentally observed net magnetization in 1.1° tDB CrI$_3$ at 0T, it requires the frustrated spin interactions due to the competition of spins between the monoclinic AFM and the rhombohedral FM regions.

- **the magnetic state with a zero total magnetization**

**Case I**: One very obvious possible configuration is that the monoclinic region has ↑↓↑↓ (or the opposite one), whereas the rhombohedral domain has ↑↓↓↑ (or the opposite one). In this configuration, each stacking region has its lowest-energy configuration, but the boundaries between them have opposite spin alignments within each of the bottom two layers. The energy penalty associated with such (intralayer) magnetic domain walls is proportional to the length of the domain boundary L multiplied by a factor of 2 because of their presence in two layers and the intralayer FM exchange coupling $J$, i.e., $E_1 \propto 2JL/a$ where $a$ is the lattice constant of CrI$_3$. Because the intralayer coupling strength is much greater than the interlayer ones, this energy penalty will become large and dominant, when the domain boundary becomes long.

**Case II**: If the energy cost for such intralayer domain boundaries becomes too high, the system will favor a different configuration, i.e., avoiding any intralayer magnetic domain boundaries. Take an extreme case example, if the area of the monoclinic domain is much larger than the rhombohedral one, we would expect ↑↓↑↓ for both monoclinic and rhombohedral stacking regions. This spin configuration does not have any intralayer domain wall penalty, although the rhombohedral one is not in its ground state and is subject to the energy penalty of $E_2 \propto |J_2| (L/a)^2$, where $L$ is the length of the domain boundary and the area shall scale with $L^2$.

Depending on the size of the domain $L$, which is controlled by the twisting angle of tDB CrI$_3$, the relative ratio between $E_1$ and $E_2$ shall change correspondingly. Most importantly, when the domain size is large (small), we shall expect $E_1 \ll E_2$ ($E_1 \gg E_2$), and thus the spin configuration shall follow the first (second) possibilities described above. Here, we emphasize again that all these spin configurations have zero total magnetization.

- **the magnetic state with a non-zero net magnetization**

**Case III**: For the intermediate domain size, where $E_1 \sim E_2$, new possible spin configuration can arise, for example, ↑↓↑↓ for the monoclinic region and ↑↑↑↓ for the rhombohedral region. In this case, the rhombohedral region is not in its ground state, because the top two layers (i.e., the top 2L CrI$_3$) both have spin up that is in contrast with the interlayer AFM coupling for a 2L CrI$_3$. This results in an energy penalty of $\sim J_1 (L/a)^2$. At the same time, the intralayer magnetic domain boundary only exists within one layer (i.e., the 2$^{nd}$ layer), which costs an energy of $\sim JL/a$. Therefore, in total, the energy cost for such a spin configuration is $E_3 \propto J_1 (L/a)^2 + JL/a$. Please note that this configuration has only one layer with intralayer magnetic domain wall (in the 2$^{nd}$ layer), so its intralayer energy cost is less than the Case I spin configuration where two layers are subject to intralayer magnetic domain wall (in the 3$^{rd}$ and 4$^{th}$ layers). This is the reason why this Case III configuration is possible to be more energetically favored than the Case I spin arrangement. In comparison to the Case II spin configuration, this Case III configuration has smaller interlayer energy cost, because $J_1 < |J_2|$.

This Case III spin configuration contains both intralayer and interlayer energy penalty ($E_3$), but the intralayer part is smaller than that of Case I ($E_1$) and the interlayer one is smaller than that of Case II ($E_2$). Therefore, when $E_1 \sim E_2$, this configuration could be most energetically favored.

- **the comparison of energies for the three spin configurations**

Below we compare the energies of these three configurations (Case I, II, and III) as a function of the domain size $L$. Here, we assume that the monoclinic region is greater, because it is the natural stacking for few-layer CrI$_3$. Here, we use $L$ to represent the linear size of the rhombohedral stacking region (normalized by the lattice constant $a$), and $L^2$ to represent its corresponding area. In reality, it should be multiplied by a geometric factor of order 1, which depends on the shape and other geometrical details of the moiré superlattices, and this factor is set to 1 here for simplicity. At a qualitative level, our conclusion is robust to the exact value of this factor. For a similar argument, we set the length of the magnetic domain boundary to be $4L$.

With this setup and the approximations, we find that the energies of the three spin configurations described above are: $E_1 \approx 8JL$, $E_2 \approx |J_2|L^2$ and $E_3 \approx 4JL + J_1L^2$. The values of $J$, $J_1$ and $J_2$ are obtained from first-principles calculations as mentioned in "background preparation" section [2,3]: $J = -2.2$ meV/$\mu_B^2$, $J_1 = 0.04$ meV/$\mu_B^2$, and $J_2 = -0.6$ meV/$\mu_B^2$.

From Fig. S1, we can see that there are two critical lengths, $L_1$ (~ $4J/J_1$) and $L_2$ as marked, at which the most energetically favored spin configuration alters. Let's discussion the three regions below, above $L_1$, between $L_1$ and $L_2$, and below $L_2$.

(a) Above $L_1$ when the rhombohedral region dominates over the monoclinic region (i.e., not quite practical – the parameter range where this simple model fails when the twist angle is so big that the inter-bilayer coupling becomes so weak), the most energetically favored spin configuration is Case I: the monoclinic region has ↑↓↑↓ (or the opposite one) and the rhombohedral domain has ↑↓↓↑ (or the opposite one). This gives zero total net magnetization.
(b) Between $L_1$ and $L_2$ when the rhombohedral and monoclinic regions are comparable (i.e., intermediate twist angle), the most energy-saving spin configuration is Case III: the monoclinic region has ↑↓↑↓ and the rhombohedral domain has ↑↑↑↓. This configuration gives a finite total net magnetization, which is consistent with our result in the 1.1° tDB CrI$_3$.
(c) Below $L_2$ when the monoclinic region dominates over the rhombohedral region (i.e., very small twist angle), the most energetically favored spin configuration is Case II: homogeneous across the monoclinic and rhombohedral regions ↑↓↑↓. This again gives zero total net magnetization.

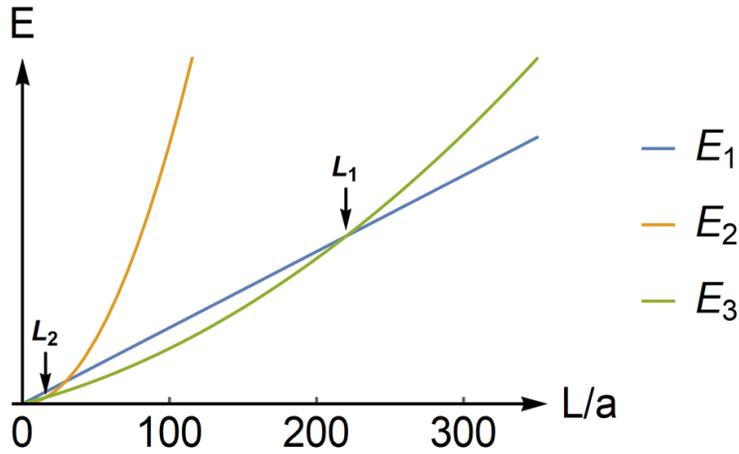

**Figure S1**  Plots of calculated energy penalty ($E_1$, $E_2$, and $E_3$) for the three spin configurations as a function of the rhombohedral stacking region size, normalized to the lattice constant $a$, i.e., $L/a$.

- **the consistency between calculations and experiment for the 1.1° tDB CrI$_3$**

We can do a quantitative comparison between our theoretical calculations and experimental observation. It should be emphasized due to the complicated nature of this material, we made approximations and simplifications in the theory treatment above. In addition, our theory has no fitting parameter at all, and the values of all control parameters are obtained from first-principles calculations. On the other hand, for small energy scales, like interlayer spin exchange, first-principles techniques are limited by resolution and accuracy, which is another source of error for our predicted values. Hence, the theory predicted values here should be treated as an order of magnitude estimation, instead of an exact solution.

With these limitations in mind, here we give key values from the theory predictions. The Case III spin configuration with a net magnetization shall arise at the critical domain size of $L_1 \approx 4J/J_1$. Using values from first-principles calculations [2,3], $L \approx 220$. This value is expected to scale with the linear size of a moiré unit cell. For our intermediate angle at which we observed Case III in the experiment, $\theta = 1.1°$, we expect $L \approx 1/\theta \approx 52$, which satisfies the requirement of $L < L_1$ and on the right order of magnitude with $L_1$.

Moreover, because the rhombohedral region has net magnetization, under an external magnetic field along (against) the direction of this net magnetization, the magnetic domain for the rhombohedral stacking shall increase (shrink) in size to reduce the energy. This magnetic domain size change is a smooth function of the magnetic field strength, and this is the reason why the signal intensity observed in this phase can vary continuously, gradually as the magnetic field increases.

## S2. Calculations of the magnetism-coupled phonon contributions in tDB CrI$_3$

In few-layer CrI$_3$, it has been established that the interlayer structural coupling can lead to Davydov splitting of the A$_g$ phonon mode of CrI$_3$, and these split phonon multiplets have two contributions in the Raman scattering: the pure structural and layered magnetism-coupled phonon contributions that correspond to the fully symmetric and antisymmetric Raman tensors [4]. Here, the layered magnetism-coupled phonon contribution encodes the static magnetic order information and tracks its evolution as a function of external magnetic field.

Here, we exploit the same idea for the strongly coupled 2L-2L regions, including both the monoclinic and rhombohedral inter-2L stacking regions, in tDB CrI$_3$. We perform the same analysis as in Ref. 4 to compute the magnetism-coupled phonon contributions for the various magnetic orders in the two stacking regions in different magnetic field ranges. The results are summarized in Table. S1 below.

Please note that the atomic displacement eigenvectors are nearly identical for the monoclinic and rhombohedral stacking regions, sharing the same symmetry properties although having slightly different atomic displacement magnitudes across the layers. Therefore, they are listed with the same sketch in Table. S1.

**Table S1.** Magnetism-phonon coupling strength ($U_i \cdot M^{M/R}$) between the phonon mode ($U_i$) and the magnetic order ($M^{M/R}$) in either monoclinic (M in the superscript, red) or rhombohedral (R in the superscript, blue) stacking regions in tDB CrI$_3$, across all four magnetic field regions with different out-of-plane spin configurations. In each entry, √ (×) stands for the presence (absence) of the magnetism-phonon coupling.

| Magnetic States | Modes | $U_1^t$ | $U_2^t$ | $U_3^t$ | $U_4^t$ |
|---|---|---|---|---|---|
| $B_\perp < B_{C1}^R$ | $M^M = \begin{bmatrix} 1 \\ -1 \\ 1 \\ -1 \end{bmatrix}$ | × | √ | × | √ |
| | $M^R = \begin{bmatrix} 1 \\ 1 \\ 1 \\ -1 \end{bmatrix}$ | √ | √ | √ | √ |
| $B_{C1}^R < B_\perp < B_{C1}^M$ | $M^M = \begin{bmatrix} 1 \\ -1 \\ 1 \\ -1 \end{bmatrix}$ | × | √ | × | √ |
| | $M^R = \begin{bmatrix} 1 \\ 1 \\ 1 \\ 1 \end{bmatrix}$ | √ | × | × | × |
| $B_{C1}^M < B_\perp < B_{C2}$ | $M^M = \begin{bmatrix} 1 \\ -1 \\ 1 \\ 1 \end{bmatrix}$ | √ | √ | √ | √ |
| | $M^R = \begin{bmatrix} 1 \\ 1 \\ 1 \\ 1 \end{bmatrix}$ | √ | × | × | × |
| $B_\perp > B_{C2}$ | $M^M = \begin{bmatrix} 1 \\ 1 \\ 1 \\ 1 \end{bmatrix}$ | √ | × | × | × |
| | $M^R = \begin{bmatrix} 1 \\ 1 \\ 1 \\ 1 \end{bmatrix}$ | √ | × | × | × |